\documentclass[aps,prb,twocolumn,amsmath,amssymb,superscriptaddress,floatfix]{revtex4-1}

\usepackage{graphicx}
\usepackage{multirow}
\usepackage{xcolor}
\usepackage{bm}
\usepackage{amsfonts,amssymb,amsmath}
\usepackage{graphicx,dcolumn,bm,color,braket,slashed}
\usepackage{times} 
\usepackage{comment}
\usepackage{array}
\usepackage{siunitx}
\renewcommand{\Vec}[1]{\mbox{\boldmath$#1$}}
\def\infinity{\infty}
\def\t#1{\textrm{#1}}
\def\ket#1{|#1\rangle }
\def\bra#1{\langle #1 |}
\def\n{\nonumber \\ }
\def\tensor{\otimes}

\newcommand\abs[1]{\left|#1\right|}

\begin{document}


\title{Two-dimensional chiral stacking orders in quasi-one-dimensional charge density waves}

\author{Sun-Woo Kim}
\affiliation{Department of Physics and Research Institute for Natural Science, Hanyang University, Seoul 04763, Korea}

\author{Hyun-Jung Kim}
\email{h.kim@fz-juelich.de}
\affiliation{Peter Gr\"unberg Institut and Institute for Advanced Simulation, Forschungszentrum J\"ulich and JARA, 52425 J\"ulich, Germany}

\author{Sangmo Cheon}
\email{sangmocheon@hanyang.ac.kr}
\affiliation{Department of Physics and Research Institute for Natural Science, Hanyang University, Seoul 04763, Korea}

\author{Tae-Hwan Kim}
\email{taehwan@postech.ac.kr}
\affiliation{Department of Physics, Pohang University of Science and Technology (POSTECH), Pohang 37673, Korea}
\affiliation{MPPHC-CPM, Max Planck POSTECH/Korea Research Initiative, Pohang 37673, Korea}

\begin{abstract}
Chirality manifests in various forms in nature.
However, there is no evidence of the chirality in one-dimensional charge density wave (CDW) systems.
Here, we have explored the chirality among quasi-one-dimensional CDW ground states
with the aid of scanning tunneling microscopy, symmetry analysis, and density functional theory calculations.
We discovered three distinct chiralities emerging in the form of two-dimensional chiral stacking orders
composed of degenerate CDW ground states: right-, left-, and nonchiral stacking orders.  
Such chiral stacking orders correspond to newly introduced chiral winding numbers.
Furthermore, we observed that these chiral stacking orders are intertwined with chiral vortices and chiral domain walls, 
which play a crucial role in engineering the chiral stacking orders.
Our findings suggest that the unexpected chiral stacking orders can open a way to investigate the chirality in CDW systems, 
which can lead to diverse phenomena such as circular dichroism depending on chirality.
%
\end{abstract}

\maketitle

\newpage

Chirality or handedness exists everywhere in nature 
and plays a significant role in all branches of the natural sciences including chemistry, biology, mathematics, and physics~\cite{Hyde1996}.
In spin- or pseudospin-ordered states, chirality manifests in various forms including magnetic chiral solitons in chiral magnets, vortices or skyrmions in thin magnetic layers, and topological monopoles in Weyl semimetals~\cite{Braun2012,nagaosa2013,armitage2018}.
Such richness is quite natural because of the vector order parameter in spin/pseudospin systems.
In contrast, it is hard to find chirality in charge-ordered states since their order parameter has a scalar nature.
A decade ago, two-dimensional (2D) $1T$-TiSe$_2$ was proposed as the first charge density wave (CDW) system with a three-dimensional (3D) real-space chiral stacking order due to inversion symmetry breaking~\cite{Ishioka2010}.
Such a chiral order of $1T$-TiSe$_2$ has been investigated in terms of chiral phase transition~\cite{Castellan2013} and optically induced gyrotropic electronic order~\cite{Xu2020}.
However, a recent sophisticated scanning tunneling microscopy (STM) investigation revealed that the intrinsic 3D chiral order (without optical induction) of $1T$-TiSe$_2$ is not allowed due to its preserved inversion symmetry between two adjacent layers~\cite{Hildebrand2018}.
Thus, the existence of the intrinsic chiral order in 2D CDW remains elusive.
Furthermore, there has been no report regarding its one-dimensional (1D) counterpart that exhibits 2D chiral orders in 1D CDW.

Recently, topological solitons with chirality are realized in quasi-1D CDW atomic wires consisting of indium (In) atoms on Si(111)~\cite{Kim2012,Cheon2015,Kim2017}.
Although the solitons are found to exhibit unusual topological properties such as $Z_4$ topology, charge fractionalization, and topological algebraic operation between them~\cite{Cheon2015,Kim2017},
their CDW ground states have not been explored in terms of chirality or chiral order. 
In this system, mirror symmetry is spontaneously broken as soon as CDW arises.
Since such broken symmetry makes CDW ground states to be geometrically chiral, this quasi-1D CDW system might show chiral CDW orders in real space. 

In this Rapid Communication, we carefully investigate the chiral order in arrays of quasi-1D CDW wires
with the aid of STM and density functional theory (DFT) calculations.
We experimentally observed three distinct 2D chiral stacking orders among CDW ground states with STM and performed extensive DFT calculations with symmetry and topology analysis to investigate their energetics as well as the microscopic mechanism behind observed chiral stacking orders.
To distinguish these chiral stacking orders, we introduced phase-shift vectors, which topologically lead to chiral winding numbers.
In addition, these 2D chiral stacking orders are intertwined with chiral vortices and chiral domain walls,
which may enable one to manipulate the emergent 2D chiral CDW orders.

The quasi-1D metallic nanowire system, In/Si(111), was grown 
by depositing one monolayer of In atoms onto the clean Si(111) surface at 700~K~\cite{Yeom1999,Kim2012}.
Subsequently, the sample was cooled down well below the ($4\times1$)--($8\times2$) CDW transition temperature of about 125~K.
STM experiments were performed in an ultrahigh vacuum (below $1\times10^{-8}$~Pa) at low temperature ($T$ = $78.150 \pm 0.001$~K). 
All STM images presented here were obtained in the constant-current mode with an electrochemically etched tungsten tip. 
To clearly visualize CDW phases, the sample bias and tunneling current were set to $-0.5$~V and 0.1~nA, respectively. 

To properly predict the energetics of the In/Si(111) system,
we have performed DFT calculations employing the Heyd-Scuseria-Ernzerhof (HSE06) hybrid functional~\cite{heyd2003hybrid,krukau2006influence} with the van der Waals (vdW) correction~\cite{tkatchenko2009accurate,zhang2011van} (referred to as HSE+vdW) within the FHI-aims code~\cite{blum2009ab}.
Note that the HSE+vdW scheme has been successfully applied to predict the energetics of $4\times1$, $4\times2$, and $8\times2$ structures as well as the band gap~\cite{kim2013driving, zhang2014stabilization, kim2015equivalence, kim2016origin}, which is consistent with previous experimental observations~\cite{tanikawa2004surface,gonzalez2009mechanism}.
Since the energy differences among various phases are small, 
we carefully performed calculations with dense 256 $k$ points per $4\times1$ unit cell and force criteria for optimizing the structures being set to $0.001$~eV/$\text{\AA}$.
The Si(111) substrate below the In wires was modeled by a six-layer slab with $\sim 30$~$\text{\AA}$ of vacuum in between the slabs. 
The bottom two-layer Si atoms with the lowest bottom Si layer passivated by H atoms were fixed during the structure relaxation.


\begin{figure}
\includegraphics[width=86mm]{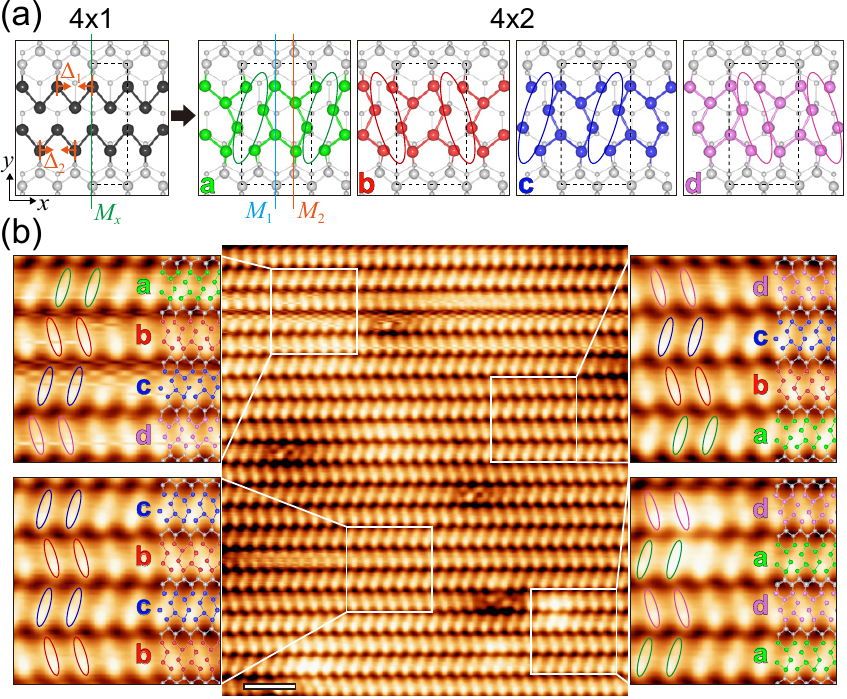} 
\caption{
(a) Atomic structures of $4\times1$ and a degenerate $4\times2$ CDW quartet ($a$, $b$, $c$, and $d$).
In atoms are represented by black and colored spheres.
Gray spheres indicate Si atoms in the zigzag chains at the surface while smaller ones do in the substrate.
Each colored oval corresponds to bright protrusions on STM images.
Vertical lines and black dashed rectangles indicate $M_x$ mirror planes and unit cells, respectively.
$\Delta_1$ and $\Delta_2$ represent the atomic dimerization displacements for two In outer subchains.
The surface coordinate system is defined by $x$ and $y$ along the $[1\bar{1}0]$ and $[11\bar{2}]$ crystallographic directions, respectively.
(b) STM image of In atomic wires grown on Si(111) obtained at 78~K. Scale bar, 3~nm.
Along the $y$ axis, there are dominant ordering patterns consisting of $4\times2$ structures in (a).
In each inset, zoom-in STM images are overlaid with the corresponding atomic configurations.
Four dominant local ordering patterns are referred to as $dcba$, $abcd$, $bc$, and $ad$ by counting from bottom to top.
}
\end{figure}

Self-assembled In nanowires on Si(111) consist of two In atomic zigzag subchains in the $[1\bar{1}0]$ direction [Fig.~1(a)], which are stitched with adjacent Si chains~\cite{Bunk1999,Gonzalez2006,Wippermann2010}.
Upon cooling, two In atomic subchains undergo a structural transition from $4\times1$ to $4\times2$ through the periodicity-doubling dimerization [$\Delta_1, \Delta_2$ in Fig.~1(a)] along a wire.
The two-way dimerization degree of freedom along both In atomic subchains spontaneously breaks the $M_x$ mirror symmetry of the $4\times1$ structure~\cite{Speiser2016}. 
This broken symmetry leads to a unique CDW quartet $\{a,b,c,d\}$, which consists of four symmetrically distinct $4\times2$ CDW ground states as shown in Fig.~1(a). 
Each $4\times2$ CDW ground state is chiral as it cannot be superposed onto its mirror image by any combination of rigid rotations and translations.
In the CDW quartet, one can further classify chiral or achiral partners depending on their symmetry relations.
For instance, $a$ is a chiral partner (mirror image) of $b$ and $d$ related by mirror operators $M_1$ and $M_2$, respectively [Fig.~1(a)].
In contrast, $a$ ($b$) and $c$ ($d$) are achiral partners to each other since they are superposed by a half-translation operator $T_x$ [see also Fig.~S1(a) in the Supplemental Material~\cite{Supple}].

Interestingly, interwire coupling in this system forces the $4\times2$ CDW quartet to exhibit unusual 2D ordering behavior perpendicular to the wires.
As shown in Fig.~1(b), each atomic wire is alternatively stacked with its chiral partners along the $y$ axis~\cite{Yeom1999}.
Such a local chiral order between two chiral partners is referred to as $8\times2$ structures in previous works.
With no preference for neighboring chiral partners, symmetrically inequivalent $8\times2$ orders are apparently intermixed as witnessed by Fig.~1(b), giving rise to $\times2$ diffraction streaks in the low-energy electron-diffraction measurements~\cite{Hatta2011}.
However, with careful examination of our low-temperature STM images, 
we found that there coexist dominant 2D chiral stacking orders among $4\times2$ building blocks [see the insets of Fig.~1(b)].
Such exotic chiral stacking orders perpendicular to the wires are unexpected and unexplored by previous works.
\begin{table}[t]
\caption{
\label{TableI}
Calculated total energies (in meV per $8\times2$ unit cell) of possible $8 \times 2$ structures
relative to the lowest energy configuration $ad$. 
}
{\renewcommand{\arraystretch}{1.7}%
\begin{tabular}{ c   c   c   c   c  }
\hline
\hline
 & $a$ & $b$ & $c$ & $d$ \\
\hline
$a$
& 84.2 & 4.3 & 106.5 & 0.0 
\\
$b$
& 4.3 & 84.2 & 0.0 & 106.5
\\
$c$
& 106.5 & 0.0 & 84.2 & 4.3
\\
$d$
& 0.0 & 106.5 & 4.3 & 84.2
\\
\hline
\hline
\end{tabular}}
\end{table}

To find the lowest energy configuration of $8\times2$,
we first calculate all possible structures constructed from a degenerate $4\times 2$ CDW quartet (Table~I). 
We perceive that only four configurations represented by $aa$, $ab$, $ac$, and $ad$ are symmetrically distinct~\cite{note}.
Other possible configurations are obtained from these four structures by applying appropriate operators such as mirror or half-translation (see Fig.~S1 in the Supplemental Material~\cite{Supple}). 
As shown in Table~I, $ad$ is the ground state stabilized over $aa$, $ab$, and $ac$ by
84.2, 4.3, and 106.5 meV per $8\times2$ unit cell, respectively.
Note that the $ab$ and $ad$ structures stacked by chiral partners are more stable than $aa$ and $ac$ stacked by achiral partners, consistent with the experimental observation in Fig.~1(b).

\begin{table}[b]
\caption{
\label{TableII}
Calculated Si dimerization magnitudes (in $\textbf{\AA}$) $\delta_1 \equiv \abs{d_2 - d_1}$ and $\delta_2 \equiv \abs{d_4 - d_3}$. Here, $d_i$ is an interatomic distance between Si atoms in the Si zigzag chains as shown in Figs.~2(a)--2(d).}
{\renewcommand{\arraystretch}{1.7}%
\begin{tabular}{ c   c   c   c   c   }
\hline
\hline
 & $aa$ & $ab$ & $ac$ & $ad$ \\
\hline
$\delta_1$
& $0.179$ & $0.167$ & $0.037$ & $0.041$
\\
$\delta_2$
& $0.185$ & $0.034$ & $0.034$ & $0.177$
\\
\hline
\hline
\end{tabular}}
\end{table}

\begin{figure}
\includegraphics[width=86mm]{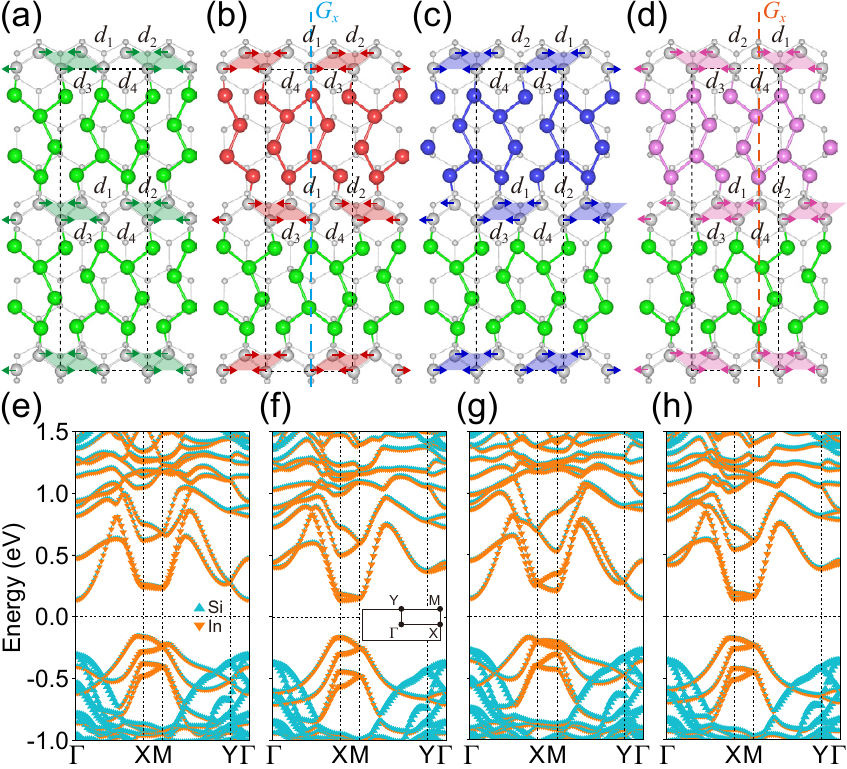} 
\caption{
(a)--(d) Four symmetrically distinct $8 \times 2$ structures ($aa$, $ab$, $ac$, and $ad$ configurations) and (e)--(h) their band structures. 
In (a)--(d), displacements of Si atoms relative to $4 \times 1$ structure (arrows) and interatomic distances between Si atoms ($d_i$; $i=1, 2, 3,$ and 4) in Si zigzag chains are indicated.
The glide mirror planes ($G_x$) are drawn in (b) and (d). 
Black dashed rectangles indicate unit cells. 
In (e)--(h), the bands projected onto orbitals of Si and In atoms are displayed where the magnitude of half-triangles is proportional to the weight. 
}
\end{figure}

To understand the mechanism for the preference of chiral partners, we compare calculated geometries and band structures for $aa$, $ab$, $ac$, and $ad$ structures (Fig.~2). 
Compared to an ideal $4\times1$ structure, each $8\times2$ structure has different dimerization patterns of Si zigzag chains [see arrows in Figs.~2(a)--2(d) and Table~II].
These dimerized Si zigzag chains mainly determine the energetics in Table~I.
Larger Si dimerization $\delta_{i=1,2}$ ($\delta_1 \equiv \abs{d_2 - d_1}$ and $\delta_2 \equiv \abs{d_4 - d_3}$, where $d_i$ is an interatomic distance between Si atoms in the Si zigzag chains) leads to larger electronic energy gain; 
thus, $aa$, $ab$, and $ad$ structures are more stable than $ac$ structure. 
However, since the cost in lattice energy is proportional to $\delta^2_i$, 
both larger dimerizations of a single Si zigzag chain ($aa$) lead to higher energy than the case of one smaller and one larger dimerization ($ab$ and $ad$). 
This finding strongly supports the alternating CDW orientations (either $ab$ or $ad$) perpendicular to the wire due to the interwire coupling. 
Moreover, we notice that the larger Si dimerizations are differently located in $ab$ and $ad$:
the larger Si dimerization $\delta_1$ ($\delta_2$) for $ab$ ($ad$) occurs right above the hollow (bonding) site of the Si substrate. 
Such a subtle difference induces the sublattice symmetry breaking, leading to the small energy difference (Table~I) between the otherwise degenerate $ab$ and $ad$ structures.
In this sense, the $ad$ configuration is a true ground state having $8\times2$ periodicity together with other symmetrically equivalent structures ($da$, $bc$, and $cb$),
as observed in Fig.~1(b).
It is noteworthy that the previous DFT studies~\cite{gonzalez2009mechanism,Wippermann2010,kim2013driving,kim2015equivalence} overlooked the difference between $ab$ and $ad$ structures: $ab$ and $ad$ were used without distinction for describing experimentally observed $8\times2$ structures. 
Here we clarify the $8\times2$ ground state and present the microscopic mechanism for considerable interwire coupling in the In/Si(111) system.

Figures 2(e)--2(h) display the calculated atom-projected band structures for $aa$, $ab$, $ac$, and $ad$ configurations.
They all show insulating electronic structures with the surface states composed of the hybridization between Si and In orbitals.
Along the $\overline{XM}$ and $\overline{Y\Gamma}$ lines, the band dispersions are not flat, which indicates that there is substantial interwire coupling in the $8\times2$ structures.
Unfavorable $aa$ and $ac$ structures have conduction band minima at the $\Gamma$ point, which disagrees with the observed insulating electronic structure showing conduction band minima at the $X$ point by time- and angle-resolved photoemission spectroscopy (trARPES)~\cite{nicholson2018beyond,nicholson2019excited}.
For $ab$ and $ad$, where the overall band structures of two configurations are nearly the same, there are twofold degeneracies along the $\overline{MY}$ line due to the glide mirror $G_x$ and time-reversal $\theta$ symmetries: Kramers-like degeneracy protected by combined antiunitary operator $(G_x\theta)^2 = -1$ along the $G_x\theta$-invariant $\overline{MY}$ line where $\theta^2 = 1$ in our spinless system~\cite{takahashi2017spinless}.
It is noteworthy that the band structure of the $ad$ configuration calculated by HSE+vdW is remarkably consistent with the trARPES experiment and is improved over previous GW calculation (see Fig.~S2 in the Supplemental Material~\cite{Supple}).


%




\begin{figure}
\includegraphics[width=86mm]{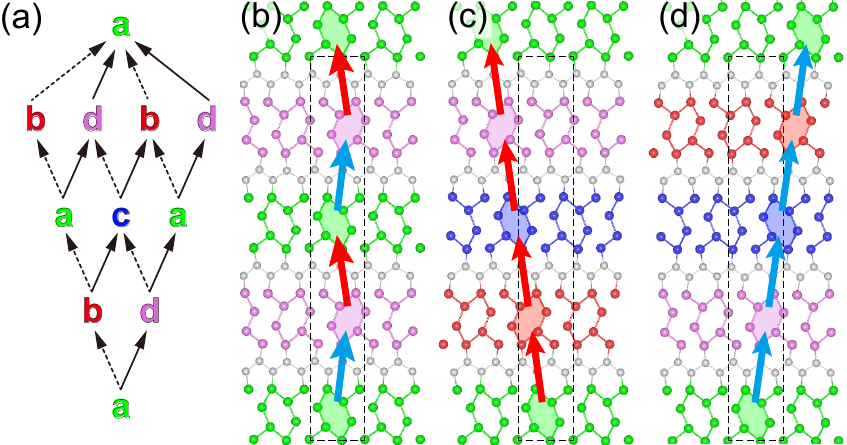} 
\caption{
(a) Stacking sequence diagram for possible $(16 \times 2)$ periodicity based on energetics in Table~I
and (b)--(d) calculated degenerate ground states with (b) nonchiral ($ad$), (c) left-chiral ($abcd$), and (d) right-chiral ($adcb$) stacking orders.
In (b)--(d), the phase-shift vectors connect one In hexagon to another nearest-neighbor In hexagon
and their colors indicate the different connecting directions.
Black dashed rectangles indicate unit cells.
}
\end{figure}

Next, we consider a longer periodicity of stacking along the $y$ axis.
To explain the observed $16\times2$ stacking periodicity in Fig.~1(b), we investigate the possible $16\times2$ ground states based on Table~I;
we present a stacking sequence diagram made out of energetically favorable building blocks ($ab$ and $ad$)  as well as their symmetric equivalence ($ba$, $dc$, $cd$; $bc$, $da$, $cb$)  [Fig.~3(a)].
Note that, since these building blocks should be composed of chiral partners, odd-periodicity stacking such as an $abc$ ($12\times2$) structure, 
which inevitably involves energetically unfavorable stacking by achiral partners ($ac$ or $ca$), is energetically not allowed.
Thus, the $16\times2$ stacking is a minimal periodicity after the $8\times2$ periodicity.

Surprisingly, our DFT calculations show 
that two $16\times2$ configurations, $abcd$ and $adcb$, are energetically degenerate with the $8\times2$ $ad$ configuration (within $\lesssim0.03$ meV per $8\times2$ unit cell), 
which nicely explains the intermixed ordering patterns in the experimental data [Figs.~1(b) and 3(b)--3(d); see also Fig.~S3 in the Supplemental Material~\cite{Supple}].
To visualize the chirality of the chiral ordering between two In atomic nanowires,
we introduce two different phase-shift vectors along the $y$ direction as indicated by red and blue arrows in Figs.~3(b)--3(d).
Using the phase-shift vectors, 2D stacking orders of the degenerate $ad$, $abcd$, and $adcb$ configurations can be geometrically distinguished.
In Fig.~3(b), the two phase-shift vectors of the $8\times2$ $ad$ structure appear alternatively along the $y$ direction, implying nonchiral stacking order.
On the other hand, the $abcd$ ($adcb$) structure shows only the left- (right-) moving phase-shift vectors along the $y$ direction, indicating its left-chiral (right-chiral) stacking order.


\begin{figure}
\includegraphics[width=86mm]{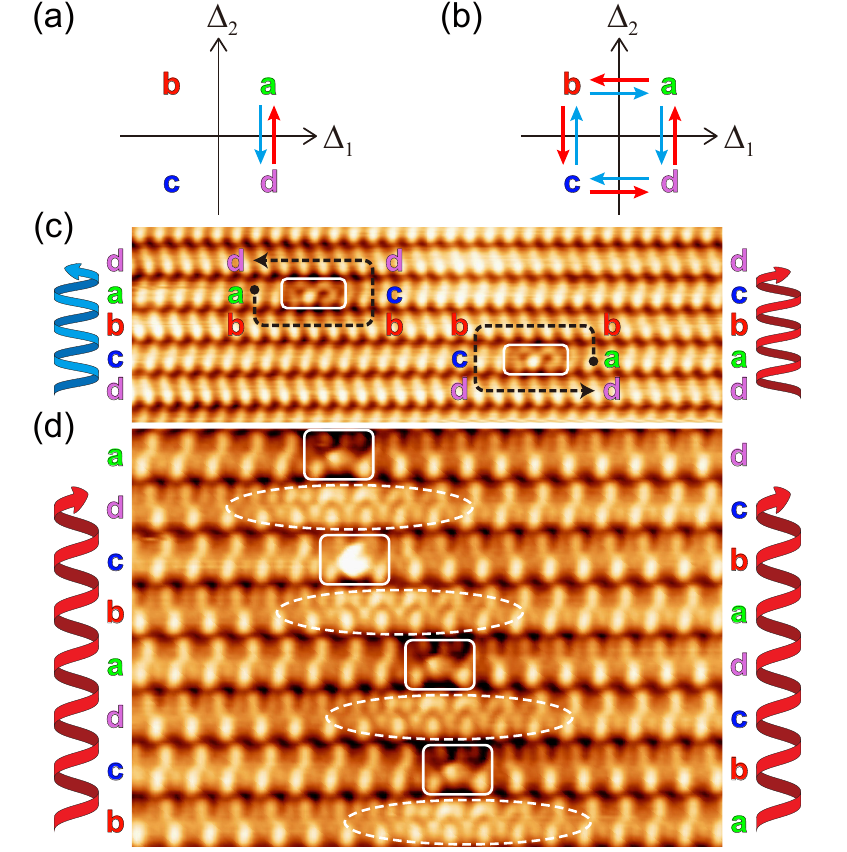} 
\caption{
(a) Nonchiral and (b) chiral stacking orders composed of two and four units of $4\times2$ structures in the parameter ($\Delta_i$) space.
Red and blue arrows correspond to the left- and right-moving phase-shift vectors along the $y$ direction, respectively.
In (b), red (blue) arrows rotate counterclockwise (clockwise), leading to the positive (negative) chiral winding number.
In sharp contrast, there is no chiral winding in the case of (a).
(c) STM image of two coexisting different chiral stacking orders and chiral vortices.   
The chiral stacking order flips over through two In-adatom defects (denoted as white rectangles),
which are chiral vortices (indicated by dashed arrows) between the opposite chiral stacking orders.
(d) STM image of a domain wall between two chiral domains with the same chiral winding number.
The domain wall consists of In-adatom defects and topological solitons, which are indicated by rectangles and ovals, respectively.
In (c) and (d), 
the helical red (blue) arrows correspond to the positive (negative) chiral winding number in (b).
}
\end{figure}

To understand topological meaning of chiral stacking orders, we characterize degenerate $ad$, $abcd$, and $adcb$ structures with their chiral winding numbers,
which are defined as the total number of turns around the gap closing point ($\Delta_1 = \Delta_2 = 0$) in the order-parameter space ($\Delta_i$).
When counting a chiral winding number, one counterclockwise (clockwise) turn corresponds to $+1$ ($-1$).
As shown in Figs.~4(a) and 4(b), all three structures uniquely show different chiral winding numbers: $N_{ad} = 0$, $N_{abcd} = +1$, $N_{adcb} = -1$.
It is noteworthy that there is a straightforward one-to-one mapping of the chiral winding numbers onto the overall phase-shift vectors described in Fig.~3.


Additionally, we discover chiral vortices and chiral domain walls that interpolate two distinct chiral stacking orders.
Chiral vortices exist between two chiral stacking orders with different nonzero chiral winding numbers [Fig.~4(c)].
The vorticity of these chiral vortices is defined by counting nearby CDW ground states from $a$, $b$, $c$ to $d$ (see Fig.~S4 in the Supplemental Material~\cite{Supple}).
A vortex (antivortex) shows a counterclockwise (clockwise) sequence as indicated by the dashed arrows in Fig.~4(c).
In contrast to a chiral vortex, a chiral domain wall, consisting of topological solitons and In-adatom defects~\cite{Song2019,Lee2019}, bridges two chiral stacking orders with the same chiral winding number [Fig.~4(d)].
These observations strongly suggest that these chiral stacking orders exhibit domain topology, which is found in 2D topological systems~\cite{Huang2017}.
As witnessed in many materials, new functionalities can be obtained by engineering chiral-ordered structures using vortices or domain walls~\cite{Huang2017}.
Since In nanowires show three distinct 2D chiral stacking orders, one can expect a new functionality such as circular dichroism,
which shows the differential absorption of left- and right-circularly polarized light~\cite{Berova2000}. 
Further study is needed to see whether one can observe and/or control the probable circular dichroism from chiral stacking orders.
For example, one may use scanning tunneling luminescence~\cite{Berndt1991,Nazin2003,Kuhnke2017}, 
which can not only measure optical response from nanometer-scale chiral stacking orders beyond diffraction limit 
but may also control chiral stacking orders by manipulating In adatoms with a scanning tip.

In summary, we found the 2D chiral stacking orders in arrays of quasi-1D CDW ground states using STM and DFT calculations.
We experimentally observed three distinct chiral stacking orders among four CDW ground states: right-, left-, and nonchiral stacking orders. 
Based on the extensive DFT calculation with symmetry and topology analysis, 
we found that the dimerized Si zigzag chain 
captures the essential physics for
the emergence of the chiral stacking orders
and classified the three chiral stacking orders by the topological chiral winding numbers.
Furthermore, topological solitons and defects play important roles as chiral domain walls and vortices between distinct 2D chiral stacking orders.
Our findings open a research platform to explore the chirality in 1D charge-ordered systems,
which may provide new functionalities such as circular dichroism.
%

\begin{acknowledgments}
This work was supported by the National Research Foundation of Korea (NRF) funded by the Ministry of Science and ICT, South Korea (Grants No. NRF-2018R1C1B6007607, No. NRF-2018R1A5A6075964, and No. 2016K1A4A4A01922028).
S.-W.K. and S.C. were supported by 
the research fund of Hanyang University (HY-2017). 
S.-W.K., H.-J.K., and S.C. acknowledge support from POSCO Science Fellowship of POSCO TJ Park Foundation.
H.-J.K. acknowledges financial support from the AIDAS project of the Forschungszentrum J\"ulich and CEA.
%
%
We thank the Korea Institute for Advanced Study for providing computing resources (KIAS Center for Advanced Computation Linux Cluster System) for this work.
\end{acknowledgments}

\bibliographystyle{apsrev4-1}

\begin{thebibliography}{40}
\bibitem [{not()}]{note}%
  \BibitemOpen
  \href@noop {} {}\bibinfo {note} {Note that the $8\times2$ $ac$ structure with
  a rectangular unit cell (Fig.~2) can be considered as the $4\times2$
  structure with a parallelogram unit cell.}\BibitemShut {Stop}%
\end{thebibliography}

\clearpage

\pagebreak

\onecolumngrid
\widetext
\renewcommand{\Vec}[1]{\mbox{\boldmath$#1$}}
\def\infinity{\infty}
\def\t#1{\textrm{#1}}
\def\ket#1{|#1\rangle }
\def\bra#1{\langle #1 |}
\def\n{\nonumber \\ }
\def\tensor{\otimes}
\setcounter{figure}{0}
\renewcommand{\theequation}{S\arabic{equation}}
\renewcommand{\thefigure}{S\arabic{figure}}

\begin{center}
\textbf{\Large{Supplemental Material: Two-dimensional chiral stacking orders in quasi-one-dimensional charge density waves
}}
\end{center}

\begin{figure}[h]
\includegraphics[width=165mm]{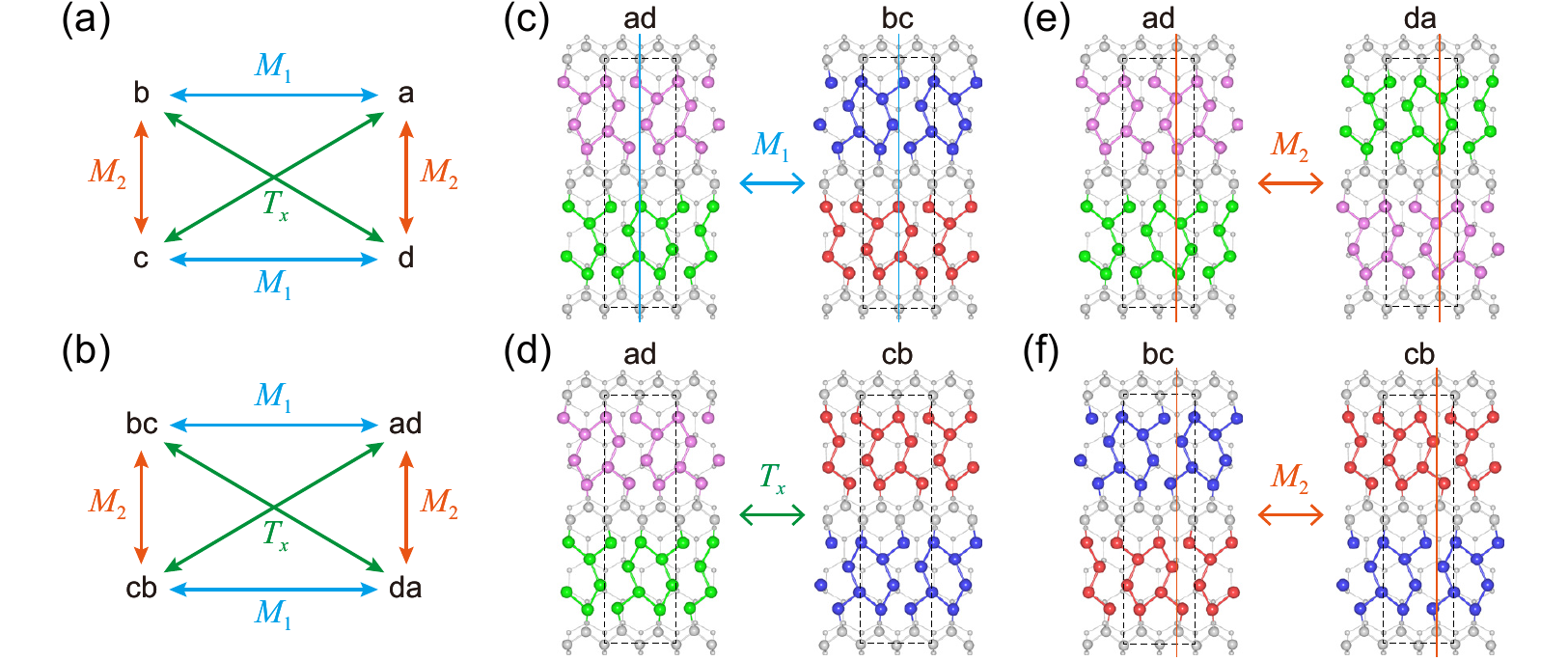} 
\caption{
Schematic diagrams for symmetry transformations among (a) four $4\times2$ ground states ($a, b, c$, and $d$) and (b) four $8\times 2$ ground states ($ad, bc, cb,$ and $da$).
Here, $M_1$ ($M_2$) is a mirror operator whose mirror plane lies at (off) the center of the unit cell as shown in (c) [(e) and (f)]. $T_x$ is a half-translation operator along the $x$ axis.
Since $ad$, $bc$, $cb$, and $da$ are related to each other by symmetry transformations, they are symmetrically equivalent configurations.
(c)--(f) Detailed examples of symmetry transformations: (c) $ad$ $\leftrightarrow$ $bc$; (d) $ad$ $\leftrightarrow$ $cb$; (e) $ad$ $\leftrightarrow$ $da$; and (f) $bc$ $\leftrightarrow$ $cb$.
Similarly, one can apply the same symmetry transformations to $aa$, $ab$, and $ac$ configurations.
In this way, we have only four symmetrically distinct $aa$, $ab$, $ac$, and $ad$ configurations out of all possible $8\times2$ configurations as discussed in the main text.
}
\end{figure}

\begin{figure}
\includegraphics[width=165mm]{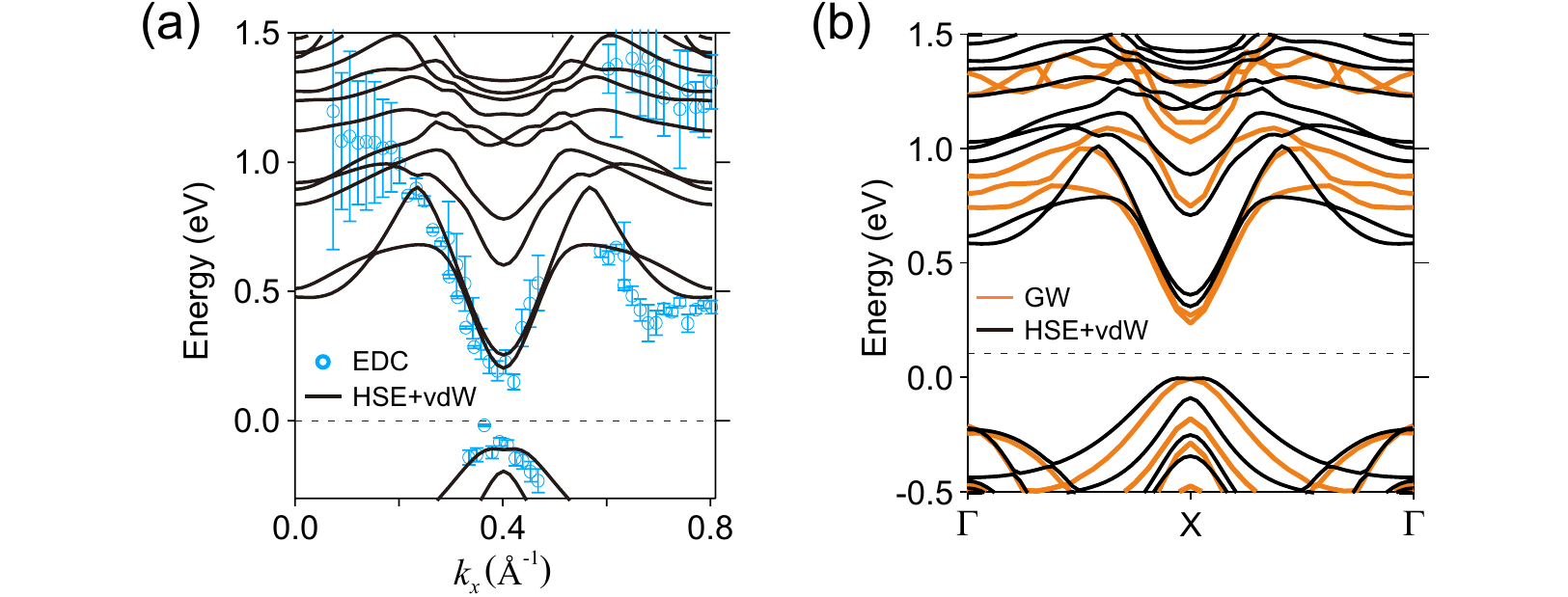} 
\caption{
(a) Comparison between HSE+vdW calculation ($ad$ structure) and energy dispersion curves obtained from time-resolved and angle-resolved photoemission spectroscopy (trARPES). 
HSE+vdW result agrees well with trARPES result.
(b) Comparison between GW and HSE+vdW calculations for $ad$ structure.
Although the HSE+vdW result is generally consistent with the previous GW calculation, HSE+vdW significantly improves the band dispersion at $\Gamma$,
which GW fails to capture. 
trARPES and GW results are taken from Ref.~32.
}
\end{figure}

\begin{figure}
\includegraphics[width=165mm]{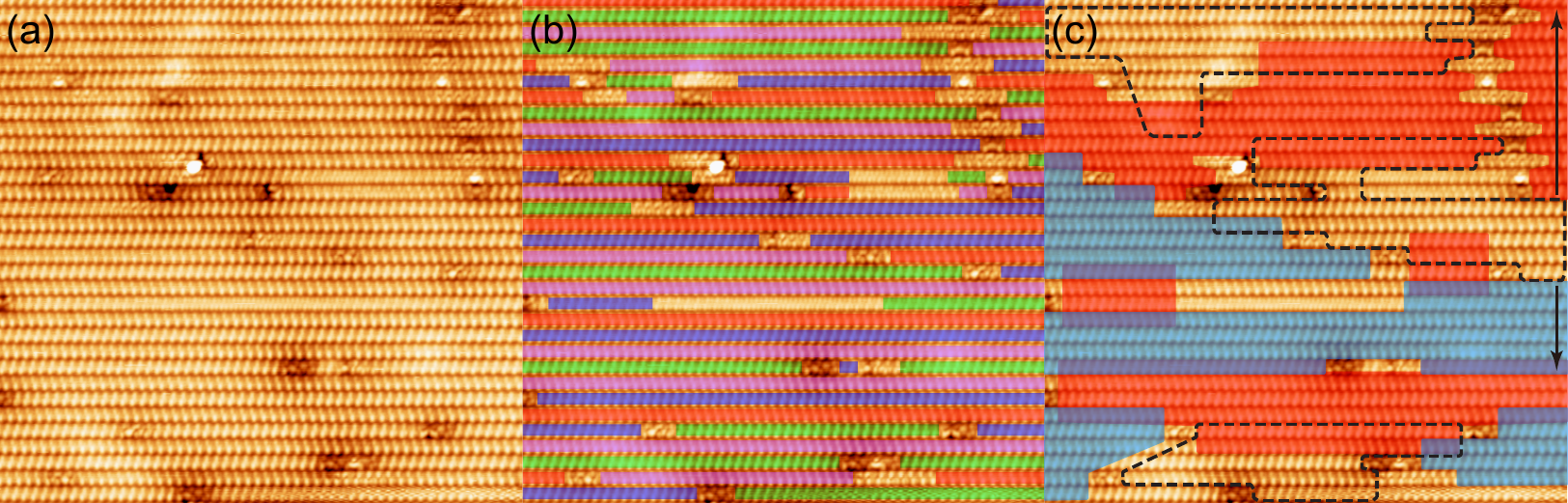} 
\caption{
(a) Large scale STM image ($40\times40$ nm$^2$) of In atomic wires with several defects and solitons.
(b) Four $4\times2$ ground states (indicated by four colors) overlaid on the same STM image in (a). 
Uncolored regions represent solitons and defects.
(c) Coexisting intertwined chiral stacking orderings overlaid on the same STM image in (a). 
Red (blue) regions indicate the left-chiral (right-chiral) stacking order of $abcd$ ($dcba$) with the positive (negative) chiral winding number 
while the regions enclosed by black dashed lines indicate the nonchiral stacking order of $ad$ (or $bc$) with the zero chiral winding number.
Black vertical arrows on the right side indicate the ascending order of four $4\times2$ ground states from $a$ to $d$.
}
\end{figure}

\begin{figure}
\includegraphics[width=80mm]{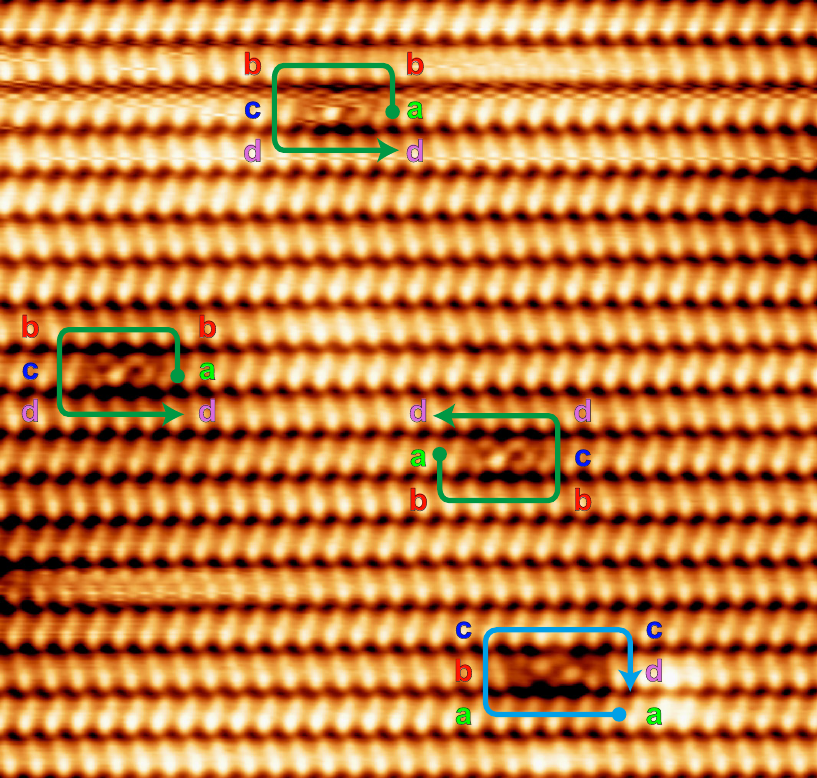} 
\caption{
STM image showing chiral vortices among different 2D chiral stacking orders.
This STM image is the same with Fig.~1(b). 
Three chiral vortices and a chiral antivortex are represented by green counterclockwise and blue clockwise arrows, respectively.
The vortex and antivortex consist of indium-adatom defects on pristine nanowires.
See more information on the indium-adatom defects in Refs.~34 and 35 of the main text.
}
\end{figure}

\end{document}